\documentclass[aps,article,author-year,notitlepage]{revtex4-1}
\usepackage{subeqn}
\usepackage{graphicx}
\usepackage{float}
\usepackage[T1]{fontenc}
\usepackage[latin1]{inputenc}
\usepackage{amssymb}
\include{epsf}
\epsfverbosetrue

\newcommand{\bea}{\begin{eqnarray}}
\newcommand{\eea}{\end{eqnarray}}
\newcommand{\Ksla}{K \!\!\!\! /}
\newcommand{\Qsla}{Q \!\!\!\! /}
\newcommand{\Psla}{P \!\!\!\! /}
\begin{document}

\large

\title{Thermal field theory at next-to-leading order in the hard thermal loop expansion}

\author{A. Mirza}
\email[]{mirzaalx@gmail.com}
\affiliation{Department of Physics, Brandon University, Brandon, Manitoba, R7A 6A9 Canada}
\affiliation{Winnipeg Institute for Theoretical Physics, Winnipeg,
Manitoba}
\author{M.E. Carrington}
\email[]{carrington@brandonu.ca}
\affiliation{Department of Physics, Brandon University, Brandon, Manitoba, R7A 6A9 Canada}
\affiliation{Winnipeg Institute for Theoretical Physics, Winnipeg,
Manitoba}

\begin{abstract}

In this paper we study the hard-thermal-loop effective theory at next-to-leading order. 
Standard power-counting predicts that a large number of diagrams, including 2-loop diagrams, may need to be calculated. 
In all of the calculations that have been done however, with the exception of the photon self-energy,  the full next-to-leading order contribution can be obtained by calculating only soft 1-loop diagrams with effective lines and vertices. It is of interest to know if the photon self-energy is the only exception to this rule, or if there are others, and which ones. In this paper we perform a refined power-counting analysis using real-time finite temperature field theory which is particularly well suited to the task.
We show that the standard power-counting rules obtained from the imaginary time formalism usually over-estimate the size of the 2-loop diagrams. We argue that the  only exceptions to the rule that the 1-loop soft diagrams give the next-to-leading order contribution are $2n$-photon vertices.

\end{abstract}
\maketitle

\section{Introduction}
\label{section:introduction}

In this paper, we discuss calculations in a relativistic statistical field theory at finite temperature. 
Using conventional notation, momenta which are of the order of the temperature ($T$) will be called `hard,' and momenta which are the order of the coupling times the temperature ($gT$) will be called `soft.'

An amplitude with at least one hard external momentum can be calculated using normal perturbation theory. 
However, if all momenta are soft, perturbation theory fails. This has been known for over 20 years, when it was first discovered that perturbative calculations of the gluon damping rate were gauge dependent. 
The solution was discovered by Braaten and Pisarski \cite{BP}. For soft momentum scales, diagrams that are formally higher order in perturbation theory contribute at lowest order.  

\begin{figure}[H]
\begin{center}
\includegraphics[width=13cm]{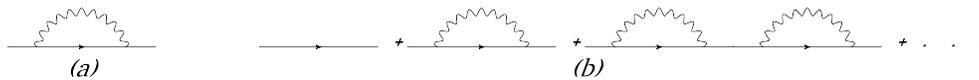}
\end{center}
\caption{Some diagrams in the HTL theory.\label{htl1}  }
\end{figure}

As an example, we consider the electron propagator. The 1-loop self-energy is shown in figure \ref{htl1}a\footnote{All figures in this paper are drawn using jaxodraw \cite{jaxo}. We use straight lines for fermions, dotted lines for ghosts, wavy lines for photons, spiral lines for gluons, and dots to indicate effective propagators and vertices.}. It is easy to show that ${\rm Tr}\,[\gamma_0 \Sigma(q_0,\vec 0)] \sim e^2T^2/q_0$, and that this result is produced by the hard region of the momentum integral. If the external momentum $q_0$ is $\sim eT$, then ${\rm Tr}\,[\gamma_0 \Sigma(q_0,\vec 0)] \sim e T$, which is the same order as the bare inverse propagator:
\bea
\label{break}
\Sigma\sim eT \sim Q \sim S_0^{-1} \,.
\eea
Since the full propagator is obtained from the Dyson equation $S^{-1}=S_0^{-1}-\Sigma$, this means that an infinite series of 1-loop self-energy insertions must be resummed to obtained the leading order (LO) propagator, for soft excitations. This result is represented in figure \ref{htl1}b.  
\par
In the same way, one finds that all bare vertices (except ghost lines and vertices with external ghost legs) are modified by 1-loop contributions that are the same order as the corresponding bare quantity. Ghost lines and vertices are not dressed. This is explained in section \ref{section:LO}. These LO 1-loop contributions to bare quantities are called hard thermal loops (HTL's).

\begin{figure}[H]
\begin{center}
\includegraphics[width=10cm]{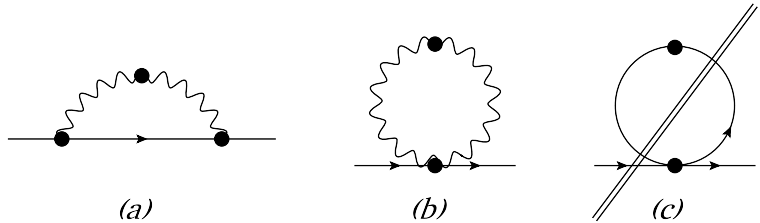}
\end{center}
\caption{Some contributions to the fermion self-energy at NLO. The first diagram is the dressed version of figure \ref{htl1}a, the second diagram does not have a bare counter-part, and the third diagram does not contribute at NLO because the HTL 4-fermion vertex is zero. \label{htl2}  }
\end{figure}

\begin{figure}[H]
\begin{center}
\includegraphics[width=7cm]{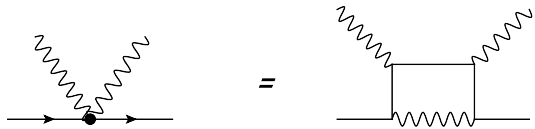}
\end{center}
\caption{An HTL 4-vertex. All legs are soft and the loop momentum is hard.  \label{htl3}  }
\end{figure}

Now we consider next-to-leading order (NLO) contributions to the fermion self-energy. The soft momentum regime in the 1-loop diagram in figure \ref{htl1}a produces a sub-leading contribution to the self-energy. In order to obtain the contribution from the soft regime of the one loop diagram, one must use effective propagators and vertices. This is shown in figure \ref{htl2}a, where resummed propagators and vertices are indicated by solid dots in the figure. 
There is an additional diagram, which does not have a bare counterpart, that contributes to the fermion self-energy at NLO. This is the tadpole diagram in figure \ref{htl2}b which appears because the effective 4-vertex is non-zero even though there is no bare 4-point vertex in QED. The HTL which produces this effective 4-vertex is shown in figure \ref{htl3}. Effective vertices with 4 or more fermion legs are zero in the HTL approximation (see section \ref{section:LO}) and for this reason, there is no tadpole diagram of the type shown in figure \ref{htl2}c in the fermion self-energy.

A reorganized perturbation theory is obtained by calculating all contributions to a given $n$-point function that can be constructed with  HTL lines and vertices. 
 The re-organized perturbation theory resumms certain infinite classes of diagrams which all contribute at NLO. It has been shown that it can be obtained from an effective action \cite{BPeffectiveaction,Taylor-Wong}, and that it produces gauge independent results \cite{KKR}.

\vspace*{.2cm}

\begin{figure}[H]
\begin{center}
\includegraphics[width=10cm]{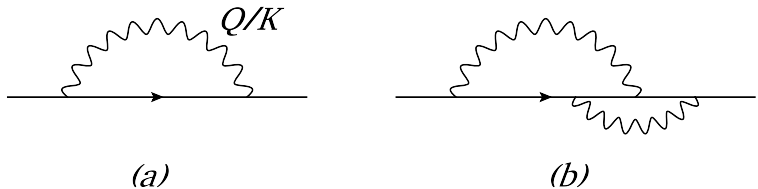}
\end{center}
\caption{Potential contributions to the self-energy at NLO. The notation $Q/K$ on the diagram in part (a) indicates that the first sub-leading term in the expansion in $Q/K$ is taken. There are three 2-loop contributions to the self-energy, one of which is shown in part b. \label{htl4}  }
\end{figure}

Standard power counting arguments using the imaginary time formalism of finite temperature field theory indicate that there are two other contributions which may be the same order as the 1-loop diagrams with soft loop momentum and resummed propagators and vertices \cite{BP}. 
Since all LO HTL's come from the hard $K$ regime of 1-loop diagrams, one extracts the HTL using an expansion in the ratio $Q/K$, and sub-leading terms in this expansion may contribute at NLO.\footnote{For tadpole diagrams there is no expansion since the tadpole self-energy is momentum independent.}
In addition, there may be contributions from 2-loop diagrams, with both loop momenta hard. 
For the example of the fermion self-energy, these two types of possible NLO contributions are shown in figures \ref{htl4}a and \ref{htl4}b (we show only one of several possible 2-loop diagrams in the figure).  We refer to the contributions in figures \ref{htl2}a and \ref{htl2}b as ``1-loop soft,'' those in figures \ref{htl4}a and \ref{htl4}b are called ``1-loop hard corrected,'' and ``2-loop hard'' contributions, respectively. 

A summary of the NLO calculations that have been done is given in section \ref{section:table}. In all but one of these calculations, the full NLO result is obtained from the (1-loop soft) piece. The  photon self-energy is the only known example  for which the entire NLO contribution is not obtained from the (1-loop soft) contribution.
Thus it seems likely that standard power-counting arguments over-estimate the contribution from (1-loop hard corrected) and (2-loop hard) terms in many cases. This issue has never been throughly investigated. In this paper we perform a more refined power-counting analysis using the Keldysh representation of real-time finite temperature field theory. We argue that the only exceptions to the rule that the full NLO result is obtained from the (1-loop soft) diagrams are $2n$-photon vertices. 

We comment that it is not unexpected that power-counting at NLO is considerably more complicated than at LO.
The reason that standard perturbation theory fails at LO in a thermal field theory is related to the existence of a new dimensional scale (the temperature).
At NLO, we have yet another dimensional scale, the thermal mass that was generated at LO by the HTL, which complicates power-counting even more. 

This paper is organized as follows. In section \ref{section:notation} we introduce our notation, and in section \ref{section:LO} we review LO power-counting. In sections \ref{section:1hc}, \ref{section:2h} and \ref{section:1s} we derive the orders of the (1-loop hard corrected), (2-loop hard) and (1-loop soft) contributions for the 2-point function. We show that for the 2-point function the (1-loop soft) contribution dominates, in all cases but the photon self-energy. In section \ref{section:photons} we explain why the photon self-energy is an exception, and show how one can understand the NLO calculation using power-counting arguments. In section \ref{section:table} we summarize in table form the results of all calculations that have been done for the LO and NLO 2-point functions, and we discuss vertex functions with more than two legs. 
In section \ref{section:conclusions} we present our conclusions.

\section{Notation}
\label{section:notation}

From the point of view of power-counting, the difference between QCD and QED is that the former contains vertices with 3 and 4 gauge bosons, and the latter has no effective $(2n+1)$-photon vertices (at any order), because of Furry's theorem. In many cases, these difference do not change the power-counting results. Accordingly, we drop colour indices and group factors and do not distinguish, for example, the contribution to the gluon self-energy from the quark bubble and the contribution to the photon self-energy from the electron bubble. In some cases, the differences between QCD and QED are important and will be emphasized. 
We work throughout in the Lorentz-Feynman  gauge. We use the following zero temperature notation for the gauge boson and fermion propagators and self energies:
\bea
&& D^{0}_{\mu\nu}(P) = -\frac{g_{\mu\nu}}{P^{2}}\,,  \\[2mm]
&& D_{\mu\nu}(P)^{-1} = D_{\mu\nu}^{(0)-1}(P) - \Pi_{\mu\nu}(P)\,, \nonumber\\[2mm]
&& S_{0}(P) = (\Psla-m+i\epsilon)^{-1} = \frac{\Psla+m}{P^{2}-m^{2}+i\epsilon} \,,\nonumber\\[2mm]
&& S^{-1}(P) = S_0^{-1}(P)-\Sigma(P)\,.\nonumber
 \label{barepropagator}
 \eea
 Using this notation, a line is $i$ times the propagator, and a contribution to the self-energy is obtained from $i$ times the corresponding diagram. In addition, we define projection operators that separate transverse and longitudinal contributions to the gauge boson propagators, and pieces of the fermion propagator that correspond to positive and negative chirality over helicity ratio. These definitions are:
 \bea
&& P^{T}_{\mu\nu} = \gamma_{\mu\nu} - \frac{\kappa_{\mu}\kappa_{\nu}}{\kappa^{2}}\,,~~ P^{L}_{\mu\nu} = g_{\mu\nu}-\frac{P_{\mu}P_{\nu}}{P^{2}} - P^{T}_{\mu\nu}\,,~~P^{0}_{\mu\nu} = \frac{P_{\mu}P_{\nu}}{P^{2}}\,, \\[2mm]
&& \gamma_{\mu\nu} = g_{\mu\nu} - U_{\mu}U_{\nu}\,,~~\kappa_{\mu} = \gamma_{\mu\nu} P^{\nu}\,,~~U_{\mu} = (1,0,0,0)\,, \nonumber\\[4mm]
&& P_{+} = \frac{1}{2}(\gamma_{0}+\vec{\gamma}\cdot\hat{p}) \,,~~ P_{-} = \frac{1}{2}(\gamma_{0}-\vec{\gamma}\cdot\hat{p}) \,.\nonumber
 \eea
 The components of the gauge boson and fermion self energies are defined:
 \bea
 \label{SEdecomp}
 && -\Pi_{\mu\nu}(P) = P^{T}_{\mu\nu}\Pi_{T}(P)+P^{L}_{\mu\nu}\Pi_{T}(P)\,, \\[2mm]
 && \Sigma(P) = \Sigma^{(0)}(P)\gamma_{0} - \Sigma^{(s)}(P)\vec{\gamma}\cdot \hat{p}  = P^+\Sigma^-(P)+P^-\Sigma^+(P) \,,\nonumber
 \eea
 from which we obtain:
 \bea
&& D_{\mu\nu}(P) = -\frac{P_{\mu\nu}^{T}}{P^{2}-\Pi_{T}(P)}-\frac{P_{\mu\nu}^{L}}{P^{2}-\Pi_{L}(P)} -\frac{P_{\mu\nu}^{0}}{P^{2}} \,, \\[2mm]
&& S(P) = \frac{1}{p_{0}+p-\Sigma^{-}(P)}P^{+}+\frac{1}{p_{0}-p-\Sigma^{+}(P)}P^{-}  =: \Delta^{-}(P)P^{+}+\Delta^{+}(P)P^{-}\,.\nonumber
\end{eqnarray}

The statistical distribution functions are defined:
\bea
\label{defn:distro}
n_b(p_0) = \frac{1}{e^{\beta p_0}-1}\,,~~n_b(p_0) = \frac{1}{e^{\beta p_0}+1}\,,\\[2mm]
N_b(p_0) = 1+2n_b(p_0)\,,~~N_f(p_0) = 1-2n_f(p_0)\,.\nonumber
\eea

 Throughout this paper we use $Q=(q_0,\vec q)$ for the soft external momentum of a self-energy diagram. For diagrams with $E$ external legs, we use $Q_i$ with $i=1,2,3,\cdots E$. For all HTL's, we label the hard loop momentum $K$. 
For a 1-loop diagram with soft loop momentum, we label the loop momentum $P$. 
We will also discuss some 2-loop diagrams, and label the two hard loop momenta $K$ and $L$. 

\section{Review of LO power-counting}
\label{section:LO}

In this section we review power-counting at LO. For an arbitrary diagram, we use the following notation:
\begin{eqnarray}
&& {\rm Number~of~loops} = m  \\
&& {\rm Number~of~vertices} = v \nonumber \\
&& {\rm Number~of~propagators} = I \nonumber \\
&& {\rm Number~of~external~legs} = E \nonumber \\
&& {\rm Number~of~external~boson~legs} = E_b \nonumber \\
&& {\rm Number~of~external~fermion~legs} = E_f \nonumber 
\label{diagramparameters}
\end{eqnarray}

First we discuss a naive (incorrect) way to determine the order of an arbitrary 1-loop diagram with soft external momenta, and explain why it does not work. 
We consider again the example of the fermion self-energy and look at the component $\Sigma^{(0)}$ (see equation (\ref{SEdecomp})). At zero temperature the Feynman integral has the form:
\bea
\Sigma_{{\bf T}=0}^{(0)}(Q) \sim e^2\int dK \frac{{\rm Tr}[\gamma^0 \gamma_\mu (\Ksla+\Qsla)\gamma^\mu]}{(K^2+i\epsilon)((K+Q)^2+i\epsilon)}\,,~~~~dK=\frac{d^4 k}{(2\pi)^4}\,.
\eea
We try to guess how this integral would behave at finite temperature if $Q\sim g T$. 
It is clear that the dominant part of the integral comes from hard loop momentum, because this maximizes the phase space. 
Since we have $Q\ll K$ we set $Q=0$ everywhere in the integrand. The zero-temperature integral is ultra-violet divergent, 
but at finite temperature integrals are weighted by distribution functions of the form ${\rm lim}_{k\to\infty}n(k) \sim e^{-k T}$, which will provide a cut-off of the order of the temperature. 
This reasoning gives that the diagram is of order $e^2 T$, a result that could be obtained directly from dimensional analysis and amounts to the conclusion that the order of the diagram is determined by the number of vertices, which is equivalent to using standard perturbation theory.
In order to see why this is incorrect, we must look at the finite temperature integrand more carefully. 

We work in the Keldysh representation of the real-time formulation of finite temperature field theory, which is particularly well suited to study power counting. Below we explain briefly the structure of the Feynman rules in the Keldysh representation. Details can be found in \cite{Fugleberg2007}. 
The propagators are $2\times 2$ matrices of the form:
\begin{eqnarray}
\label{keldyshProp1}
&& D_{\mu\nu}(K) = -g_{\mu\nu} {\bf \Delta}_b(K) \,,~~~~~ S(K) = (\displaystyle{\not}K + m) {\bf \Delta}_f(K)\,,\\[2mm]
\label{keldyshProp2}
&&	{\bf \Delta}_x(K)=\left(\begin{array}{cc} \Delta_{{\rm s}x} & \Delta_{\rm r} \\ \Delta_{\rm a} & 0 \end{array}\right)\,,~~~~~x\in\{b,f\}\,,\nonumber\\[2mm]
\label{keldyshProp3}
&&	\Delta_{{\rm s}b/{\rm s}f}=N_{b/f}(p_0)\big(\Delta_{\rm r}-\Delta_{\rm a})\,,~~~~~\big(\Delta_{\rm r}-\Delta_{\rm a})=-i\,\pi\, {\rm Sgn}(k_0)\,\delta(K^2)\,, \nonumber
\end{eqnarray}
where $\Delta_{\rm r}$ and $\Delta_{\rm a}$ are the retarded and advanced scalar propagators. Vertices are temperature independent, but they also have a tensor structure: 3-vertices are $2\times 2\times 2$ tensors, 4-vertices are $2\times 2\times 2\times 2$ tensors, etc. The bare 3-point vertex can be written:
\bea
{\bf \Gamma}^0 &&  = \{\{\{g_{111},g_{112}\},\{g_{121},g_{122}\}\},\{\{g_{211},g_{212}\},\{g_{221},g_{222}\}\}\} \,,\\
&&  = \{\{\{0,-i\},\{-i,0\}\},\{\{-i,0\},\{0,-i\}\}\}\,. \nonumber
\eea
The corresponding expression for the 4-vertex is given in \cite{Fugleberg2007}.

Using these Feynman rules it is straightforward to show that each term in the integrand of a 1-loop diagram contains one symmetric propagator. The integrand for the retarded fermion self-energy has the form:
\bea
\Sigma^{(0)}(Q) \sim e^2\int dK \frac{{\rm Tr}[\gamma^0 \gamma_\mu (\Ksla+\Qsla)\gamma^\mu]}{(K+Q)^2 +i {\rm Sgn}(k_0+q_0)\epsilon }N_b(k_0){\rm Sgn}(k_0)\delta(K^2)~+~\cdots
\eea
where the dots indicate a second term which has the same structure as the first, but contains a fermion distribution function. We suppress the subscript which indicates the retarded component of the self-energy, and throughout this paper we use the convention that a finite temperature self-energy written without a subscript to indicate the component is always taken to be retarded. 
The delta function comes from the symmetric propagator (see equation (\ref{keldyshProp1})). Due to the presence of this delta function, the non-symmetric propagator contributes $1/(K+Q)^2\sim 1/(K\cdot Q)$ to the integrand. Since $Q$ is soft by assumption, we obtain a factor $1/e$ which increases the order of the HTL relative to the naive dimensional estimate. Thus we find  $\Sigma\sim eT^2/Q \sim eT$ which means that the self-energy is the same order as the inverse propagator (see equation (\ref{break})), and ordinary perturbation theory fails.

It is easy to generalize this discussion to an arbitrary 1-loop retarded $n$-point function. To obtain a specific expression for the integrand, we would need to specify which is the leg with respect to which the $n$-point function is retarded, but in order to obtain a power-counting estimate we do not need to choose a specific retarded vertex function. 

\begin{figure}[H]
\begin{center}
\includegraphics[width=15cm]{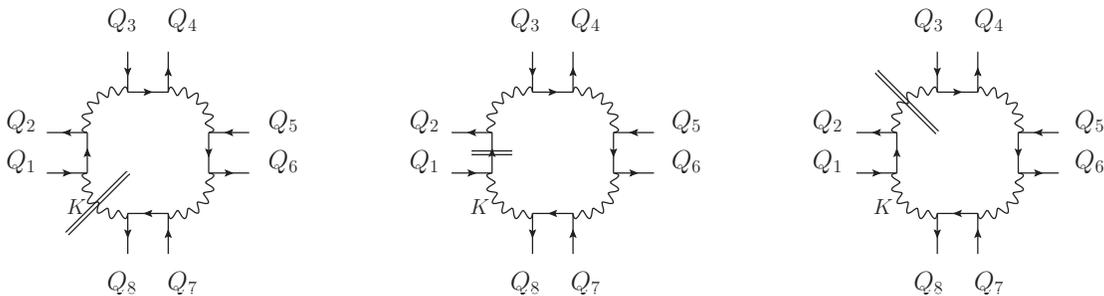}
\end{center}
\caption{Three of the eight terms in the integrand for a diagram with 8 external fermions. The symmetric propagators are marked with a double line. \label{symAll}  }
\end{figure}

We first consider 1-loop diagrams with an arbitrary number of legs, at least two of which must be fermions. The case $E_f=0$ is special and will be dealt with separately. Using Keldysh propagators and vertices and performing all tensor contractions we obtain a set of terms each of which contains one symmetric propagator, $I-1$ retarded or advanced propagators and $I=v$ vertices. The symmetric propagator can be on any one of the internal lines, and thus there are $I$ terms in the sum. As an example, the first three terms in the integrand for the 1-loop diagram with 8 external fermions are shown in figure \ref{symAll}.
Consider the first diagram in figure \ref{symAll}. The symmetric propagator contains a factor $\delta(K^2)$  which means that all of the non-symmetric propagators contribute to the integrand a factor of the form:
\bea
\Delta(K+\sum Q_j) \approx \frac{1}{K^2+2K\cdot\sum Q_j} = \frac{1}{2K\cdot\sum Q_j}\sim \frac{1}{g T^2}\,,
\eea
where we have dropped all terms proportional to $Q^2$ since we have $K\gg Q$, and $\sum Q_j$ is a symbolic notation that indicates the sum of some set of external momenta.
It is clear that for the other two diagrams in figure \ref{symAll}, and every other term in the integrand, we can always shift the loop momenta so that the symmetric propagator carries momentum $K$, and the non-symmetric propagators have arguments of the form $K+\sum Q_j-\sum Q_{j^\prime}$. 
The conclusion is that each term in the integrand contains $I-1$ non-symmetric propagators each of which contributes a factor of order $\sim 1/(gT^2)$. 
The order of a 1-loop diagram with soft external legs ($E_f\ne 0$) and dimension $d$ is therefore:
\bea
\label{LObasic}
\Gamma(E,E_f\ne 0)\sim g^v T^{d}\cdot\frac{T^{I-1}}{{Q^{I-1}}}\,,
\eea
where $g^v T^{d}$ is what one would obtain from ordinary perturbation theory, and the second factor is produced by the non-symmetric  propagators.  

It is useful to rewrite equation (\ref{LObasic}) in terms of the external variables. For a 1-loop diagram we have $v=I$ and dimensional analysis gives $d=4-E_b-\frac{3}{2}E_f$. Using $Q\sim gT$ we obtain:
\bea
\Gamma(E,E_f\ne 0)\sim g T^{4-E-E_f/2}\,.
\eea

\begin{figure}[H]
\begin{center}
\includegraphics[width=4cm]{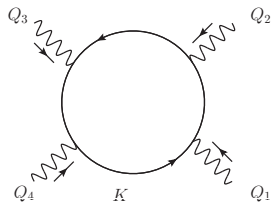}
\end{center}
\caption{An example of a diagram that has a statistical factor. \label{fig:statFac}  }
\end{figure}

For $E_f= 0$ (and $I\ge 2$) there is an additional suppression factor which comes from the thermal distribution functions. A derivation of this result for diagrams with an arbitrary number of external legs can be found in \cite{AM-thesis}. The basic idea is that the $I$ terms that are produced by the tensor contraction have the same form, except that they contain distribution functions whose momentum arguments differ by soft momenta. Combining terms one finds that if all distribution functions are either boson or fermion distributions, the lowest order term cancels and one obtains an extra factor $q_0/T$. As an example, an amplitude with four photon legs and an internal fermion loop is shown in figure \ref{fig:statFac}. The integrand will contain thermal factors of the form $\big[n_f(k_0)-n_f(k_0+q_{1\,0})\big]$, $\big[n_f(k_0+q_{1\,0})-n_f(k_0+(q_{1\,0}+q_{2\,0})\big]$, $\cdots$ which combine to produce an extra factor of order $Q/K$. 

We refer to this extra factor as the `statistical factor.' 
It is only possible to produce a one loop diagram in which all fields in the loop have the same statistics if $E_f$=0. Therefore we can write a general result that includes the statistical factor using a Kronecker delta function:
\bea
\label{LOorder}
\Gamma(E,E_f)\sim g^{\delta_{E_f\,0}} \cdot g T^{4-E-E_f/2}\,.
\eea
Dividing by $(gT)^{4-E-E_f/2}$ we obtain a result for the order of a dimensionless soft amplitude:
\bea
{\cal O}(E,E_f)\sim \big[g^{\delta_{E_f\,0}} \,g^{E_f/2-1}\big]\,g^{E-2}\,.
\eea
The factor in square brackets is one if $E_f=0$ or 2 but smaller than one for larger values of $E_f$, and therefore HTL's are zero for $E_f\ge 4$. 

\begin{figure}[H]
\begin{center}
\includegraphics[width=7cm]{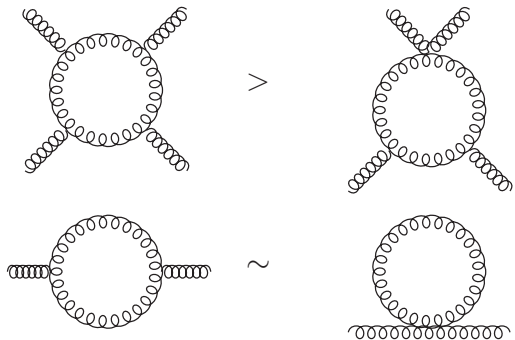}
\end{center}
\caption{Diagrams which contain 4-vertices are suppressed relative to those that do not, except for tadpole diagrams. \label{fig:htl5}  }
\end{figure}

In cases where more than one diagram contributes to the self-energy, it is not necessarily true that all diagrams contribute at leading order. 
With the exception of tadpoles, all diagrams with 4-gluon vertices are suppressed relative to diagrams with only 3-gluon vertices. The basic reason is that replacing two 3-gluon vertices with a 4-gluon vertex removes two factors $\sim K$ (from the 3-vertices) and one propagator $\sim$ $1/(K\cdot Q)$, which reduces the order of the diagram by one power of the coupling. Tadpoles are an exception to this rule, they contribute at the same order as graphs which contain 3-vertices. The reason is that they contain only one propagator which makes it impossible to obtain a statistical suppression factor, and therefore the loss of the factor $1/(K\cdot Q)$ introduced by the 4-vertex is compensated for by the lack of a statistical suppression factor. The result is that  (\ref{LOorder}) gives the correct LO for all diagrams including tadpoles, but for tadpoles the factor $g^{\delta_{E_f\,0}}$ cannot be interpreted as a statistical suppression factor. The role of 4-point vertices in LO HTL is represented symbolically in figure \ref{fig:htl5}. 

All diagrams with ghost external lines do not have HTL's. This is a simple consequence of the structure of the ghost vertex, which depends only on the momentum of the incoming ghost. Every diagram that contains an external ghost line will have an extra power of $g$ from the vertex attached to the incoming ghost. 

On physical grounds, we are especially interested in the self-energies. The real part of the resummed 2-point function has a simple physical interpretation in terms of collective modes, and the imaginary part can be used to obtain rates for scattering and production processes. In section \ref{section:NLO2} we will consider the NLO 2-point function with zero 3-momentum ($\vec q=0$). We will want to compare the order of the NLO contributions with the corresponding LO result as given by equation (\ref{LOorder}). However, (\ref{LOorder}) does not take into account the fact that the HTL can be zero due to kinematic constraints. 
It is easy to see that the imaginary part of a bubble diagram is zero, not only for $\vec q=0$ but for any time-like excitation ($Q^2>0$).
The imaginary part can be written as the square of an amplitude at both zero and finite temperature \cite{Carrington2002} and therefore contains an additional on-shell propagator, relative to the real part. For a bubble diagram, the two on-shell propagators contribute factors:
\bea
\label{kinematic}
\int dK \delta(K^2)\delta\big((K+Q)^2\big) \sim \int d k\,k^2\int_{-1}^1 dx\frac{1}{2k}\delta\big(k(q_0 -q x)\big)~~{\rm for}~~\vec k\cdot \vec q=k q x\,.\nonumber\\
\eea
The delta function has support only for $Q^2<0$ and therefore the integral is zero for time-like excitations. The tadpole is pure real and thus the imaginary part of the HTL self energy is identically zero in the limit $\vec q=0$. 

Combining these results, the leading order HTL self-energies are:
\bea
\label{LOres}
\begin{array}{ll}
{\rm Re}\Sigma_{\rm LO}\sim g T\,,~~~~~ & {\rm Re}\Pi_{\rm LO}\sim g^2 T^2\,,\\[2pt]
{\rm Im}\Sigma_{\rm LO}\sim 0\,,~~~~~ & {\rm Im}\Pi_{\rm LO}\sim 0\,.
\end{array}
\eea

\section{NLO 2-point functions}
\label{section:NLO2}

We discuss below the three potential contributions at NLO which were identified in Ref. \cite{BP}.  As discussed in section \ref{section:introduction}, we call these (1-loop hard corrected), (2-loop hard), and (1-loop-soft) contributions. 
The arguments we present in this section give the order of the diagram up to possible logarithmic factors, which will be discussed in section \ref{section:photons}. 
Our purpose is only to compare the relative sizes of the three potential contributions at NLO and we will show that in most cases it is not necessary to consider logarithmic factors to do this. The exception is the  photon self-energy. In all other cases  the full NLO contribution can be obtained from the (1-loop soft) piece which is larger than the (1-loop hard corrected) and (2-loop hard) contributions by at least one power of the coupling. This conclusion is not affected by possible logarithmic factors.

From the spectral representation, one can show that the real and imaginary parts of the self energies have definite parity under the transformation $q_0\to-q_0$. We consider only the components of the photon self-energy which are directly related to the screening mass, plasma frequencies and damping rates, and rates for photon and di-lepton production. Denoting $\Pi_{\rm L}$, $\Pi_{\rm T}$, $\Pi_{00}$ and $\Pi_\mu^{~\mu}$ generically by $\Pi$ we have:
\bea
\label{parity}
\begin{array}{ll}
{\rm Re}\,\Pi(q_0,q) = {\rm Re}\,\Pi(-q_0,q)\,,~~~~& {\rm Im}\,\Pi(q_0,q) = -{\rm Im}\,\Pi(-q_0,q)\,,\\[2mm]
{\rm Re}\,\Sigma^{(0)}(q_0,q) = -{\rm Re}\,\Sigma^{(0)}(-q_0,q)\,,~~~~& {\rm Im}\,\Sigma^{(0)}(q_0,q) = {\rm Im}\,\Sigma^{(0)}(-q_0,q)\,,\\[2mm]
{\rm Re}\,\Sigma^{(s)}(q_0,q) = {\rm Re}\,\Sigma^{(s)}(-q_0,q)\,,~~~~&{\rm Im}\,\Sigma^{(s)}(q_0,q) =- {\rm Im}\,\Sigma^{(s)}(-q_0,q)\,.
\end{array}
\eea
This result can be used to determine the order of (1-loop hard corrected) and (2-loop hard) contributions to the NLO self-energies. For (1-loop soft) contributions there is a new dimensionful scale, the thermal mass, and the parity of the self-energy is not useful.

\subsection{1-loop-hard-corrected}
\label{section:1hc}

As discussed in section \ref{section:LO}, the imaginary part of the HTL is identically zero for time-like excitations. Keeping the next order in the $Q/K$ expansion will give a non-zero contribution, but at NLO the delta function in equation (\ref{kinematic}) becomes $\delta\big[(K+Q)^2\big]\sim \delta\big(2k q_0+q_0^2 \big)\sim \frac{1}{2q_0}\delta(k+\frac{q_0}{2})$
and thus always restricts $K$ to the soft part of the phase space, which means that the integral gives a piece that is already included in the (1-loop soft) part. 
For the real part, it is easy to see that the (1-loop hard corrected) contributions are of order:
\bea
\label{1loopCorrecteda}
{\rm NLO} \sim \left(\frac{q_0}{T}\right)^2\,{\rm LO} \sim g^2 \,({\rm LO})\,.
\eea
The term in the expansion with one power of $q_0/T$ cancels identically, which can be seen without calculation directly from (\ref{parity}). From (\ref{LOres}) and (\ref{1loopCorrecteda}) we obtain:
\bea
\label{1loopCorrected}
{\rm 1\!\!- \!\!loop~ hard ~corrected:}&&~~~ {\rm Re}\Sigma_{\rm 1lhc} \sim g^2\,{\rm (LO)} \sim g^3 \, T \,,~~~\\
&& ~~~{\rm Re}\Pi_{\rm 1lhc} \sim g^2\,{\rm (LO)} \sim  g^4 \,T^2\,.\nonumber
\eea

\subsection{2-loop hard}
\label{section:2h}

There are two possible ways to construct a 2-loop diagram.

\begin{figure}[H]
\begin{center}
\includegraphics[width=12cm]{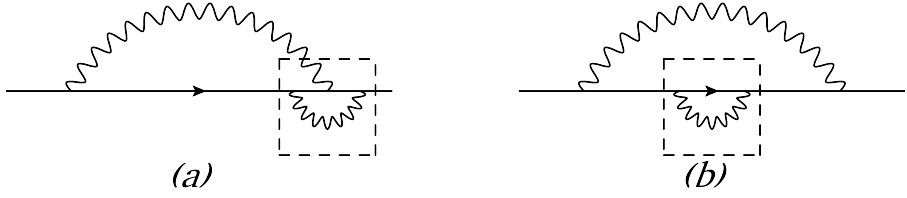}
\end{center}
\caption{There are two ways to construct a 2-loop diagram. Vertex and propagator insertions are indicated with a dotted box.  \label{2loopEx}  }
\end{figure}

(1) One can start from a 1-loop diagram and draw an additional line that joins the loop on either side of at least one external leg. For the fermion self-energy diagram, this is shown in figure \ref{2loopEx}a. This type of graph is called a vertex correction graph, because it can be thought of as a vertex corrected 1-loop graph. This is indicated by the dotted box in the figure. Since the vertex insertion contains one soft and two hard external legs (instead of all soft legs), the results of section \ref{section:LO} tell us that the corrected vertex is not an HTL, and therefore of order $g^3$. Thus we see immediately that all vertex correction graphs are suppressed relative to the 1-loop graph they were built from by a factor of $g^2$.

(2) One can also construct a 2-loop graph from a 1-loop graph by drawing an additional line that does not separate any of the external legs. These graphs are called propagator correction graphs (see figure \ref{2loopEx}b). Even though these graphs also contain two additional vertices, there is a simple power-counting argument that seems to indicate they are suppressed relative to the LO contribution by only one power of the coupling. We give this argument below and explain why it is wrong and the correct suppression factor is $g^2$, as for vertex graphs. 

\begin{figure}[H]
\begin{center}
\includegraphics[width=12cm]{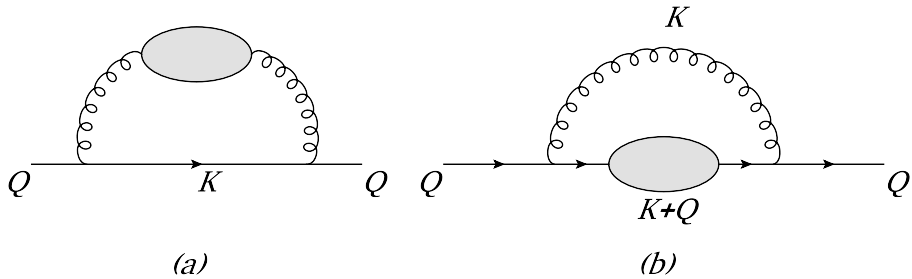}
\end{center}
\caption{2-loop diagrams contributing to the fermion self-energy that have the form of a propagator corrected 1-loop graph. The grey blob in the first diagram indicates the hard gluon self-energy, and in the second diagram it is the hard fermion self-energy. \label{fermion2loop}  }
\end{figure}

\begin{figure}[H]
\begin{center}
\includegraphics[width=12cm]{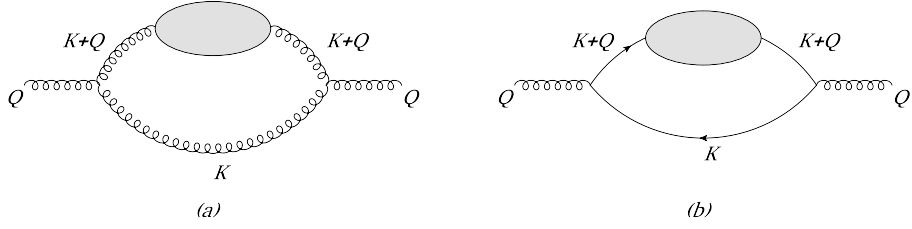}
\end{center}
\caption{2-loop diagrams contributing to the gluon self-energy that have the form of a propagator corrected 1-loop graph. The grey blobs indicate the hard gluon and fermion self-energies.  \label{gluon2loop}  }
\end{figure}

Our strategy is to combine sets of propagator corrected graphs using a hard self-energy insertion on one line, which will be represented by a grey blob. All possible propagator corrected graphs can be written as in figures \ref{fermion2loop} and \ref{gluon2loop} (the propagator corrected tadpole is not included since it is momentum independent and therefore clearly of order $g^2({\rm LO})$). The integrands are calculated in the Keldysh representation by summing over the tensor indices using the method of reference \cite{Fugleberg2007}.  

Using dimensional analysis the hard insertions are of order:
\bea
\label{hardorder}
\Pi_{\rm hard} ~\sim ~ g^2T^2 \,,~~~~ \Sigma_{\rm hard} ~\sim ~ g^2T \,.
\eea
We can try to use (\ref{hardorder}) to estimate the order of a (2-loop hard) diagram by simply dropping all factors $Q$  in the numerator and using $\delta(K^2)\Delta(K+Q)\to \delta(K^2)/(k_0q_0)$ in the denominator. We will call this the ``naive approximation'' and we will show that in all cases it gives that the 2-loop diagram is of order $g({\rm LO})$ but has the wrong parity to satisfy  (\ref{parity}). Closer inspection of the integrand reveals that it is odd in $k_0$ and therefore gives zero when integrated. This means that the  LO term is not produced by the naive approximation but contains an additional factor $q_0/k_0$, and is therefore of order $g^2({\rm LO})$ and has the right $q_0$ parity.

We start with the graph in figure \ref{fermion2loop}a. 
Using the notation of equation (\ref{keldyshProp1}) and writing explicitly the indices r and a to indicate retarded and advanced components, the integral has the form:
\bea
\label{integrand1}
\Sigma_{\rm r}^{(0)}(q_0,0)~ \,\sim \,g^2\,i\int dK\;&& {\rm Tr}\,\big[\gamma_0\gamma^\mu\Ksla\gamma^\nu\big](N_f(k_0)-N_b(k_0-q_0)))\Delta_{{\rm r}}(K)\\
&& \cdot\;\Delta_{\rm a}(K-Q)    (\Pi_{\rm hard}^{\mu\nu}(K-Q))_{\rm a}\,\Delta_{\rm a}(K-Q)\,.\nonumber
\eea
We can do the $k_0$ integration by extending to the complex plane and closing the contour in the lower half plane, so that only the poles of the retarded propagator contribute. To obtain a power-counting estimate of the real part of the integral using the naive approximation we make the replacements:
\bea
&& \Delta_{\rm r}(K) ~\to~ i\,{\rm Sgn}\,(k_0)\,\delta(K^2) \,, \\
&& \Delta_{\rm a}(K-Q) (\Pi_{\rm hard}^{\lambda\sigma}(K-Q))_{\rm a}\Delta_{\rm a}(K-Q) ~\to~ {\rm Re}\,\Pi_{\rm hard}^{\mu\nu}(K) \left(\frac{1}{k_0 q_0}\right)^2\,, \nonumber\\
&& N_f(k_0)-N_b(k_0-q_0) \to N_f(k_0)-N_b(k_0)\,.\nonumber
\eea
We obtain:
\bea
\label{se1}
{\rm Re}\,\Sigma^{(0)}(q_0,0) \,\sim \,g^2\int dK\;&& {\rm Tr}\,\big[\gamma_0\gamma_\mu\Ksla\gamma_\nu\big]\;(N_f(k_0)-N_b(k_0))\cdot\; {\rm Sgn}(k_0)\delta(K^2)  \left(\frac{1}{k_0q_0}\right)^2  {\rm Re}\,\Pi_{\rm hard}^{\mu\nu}(K)\,,\nonumber
\eea
which appears to be of order $g^2 T = g({\rm LO})$ (using $q_0\sim gT$ and  (\ref{hardorder})). 
However, the integral is even in $q_0$ and therefore has the wrong parity to satisfy (\ref{parity}).
To understand this we note that the trace contains terms proportional to
$
K_\mu g_{\nu 0}$, $K_\nu g_{\mu 0}$ and $k_0g_{\mu\nu}$. 
The first two terms give zero since $\Pi_{\rm hard}^{\mu\nu}(K)$ is transverse by the Ward identity, and the last term produces an integrand that is odd in $k_0$ which gives zero when integrated over $k_0$ (to see this remember that $N_f(k_0)$ and $N_b(k_0)$ are  odd and Re$\Pi_\mu^{~~\mu}(K)$ is even using (\ref{parity})). Thus the naive approximation is not correct, and we need to keep the first sub-leading term in the expansion in $Q/K$. This introduces another factor $q_0/k_0\sim g$ and we obtain that the integral is $\sim g^2\,({\rm LO})$ and satisfies (\ref{parity}).

The integrand corresponding to the graph in figure \ref{fermion2loop}b has the form:
 \bea
 \Sigma_{\rm r}^{(0)}(q_0,0)\,\sim \,&& g^2\,i\,\int dK\;{\rm trce}\;\,(N_f(k_0+q_0)-N_b(k_0))\Delta_{{\rm a}}(K) \Delta_{{\rm r}}^2(K+Q)
\,,\nonumber\\
&&
 {\rm trce} = {\rm Tr}\,\big(\gamma_0\gamma_\mu(\Ksla+\Qsla)\;(\Sigma_{\rm hard}(K+Q))_{\rm r}\,(\Ksla+\Qsla) \,\gamma^{\mu} \big)\,.\nonumber
 \eea
 We do the $k_0$ integration in the upper half plane picking up contributions from the poles of the advanced propagator. Making the replacements:
 \bea
&& \Delta_{\rm a}(K) ~\to~ i\,{\rm Sgn}\,(k_0)\,\delta(K^2) \,, \\
&& \Delta^2_{\rm r}(K+Q)  ~\to~ \left(\frac{1}{k_0 q_0}\right)^2\,, \nonumber\\
&& N_f(k_0+q_0)-N_b(k_0) \to N_f(k_0)-N_b(k_0)\,,\nonumber\\[2mm]
&&{\rm trce} ~\to~ {\rm trce}\bigg|_{Q=0}\,,\nonumber
\eea
we obtain:
\bea
 \label{se2}
{\rm Re}\Sigma^{(0)}(q_0,0)  \sim g^2\int dK  \,{\rm Tr}\,\big[\gamma_0\gamma_\mu\Ksla\;{\rm Re}\Sigma_{\rm hard}(k_0,k)\,\Ksla\, \gamma^\mu\big]\cdot\;\big(N_f(k_0)-N_b(k_0)\big)\;{\rm Sgn}(k_0)\,\delta(K^2)\left(\frac{1}{k_0q_0}\right)^2 \,.\nonumber
 \eea
The integral is of order $g\,({\rm LO})$ but it has the wrong parity to satisfy (\ref{parity}). Using $\Sigma(k_0,k)\sim \gamma_0 \Sigma^{(0)}(k_0,k)$ the trace gives terms proportional to $k_0^2$ and $K^2$. The factor $K^2$ is zero using the delta function, and since $N_b(k_0)$ and $\Sigma^{(0)}(k_0,k)$ are both odd the remaining term is odd and gives zero when integrated. The term containing $\Sigma^{(s)}(k_0,k)$ can be treated the same way. One obtains an additional factor $k^2/(k k_0)$ relative to the term with $\Sigma^{(0)}(k_0,k)$, and since $\Sigma^{(s)}(k_0,k)$ has opposite parity, the term of order $g({\rm LO})$ cancels. Keeping the first sub-leading term in the expansion in $Q/K$,  we obtain again that the integral is $\sim g^2\,({\rm LO})$ and satisfies (\ref{parity}).

The integrand for the graph in figure \ref{gluon2loop}a has the form:
\bea
\label{integrand1y}
\Pi^{\rm r}_{\mu\nu}(q_0,0) \sim \,g^2\,i\,\int dK\;&&{\rm vtex}_{\mu\lambda\sigma\nu}\; \bigg[\bigg(N_b(k_0+q_0)-N_b(k_0)\bigg) \\
&&\cdot\;\Delta_{\rm a}(K)\Delta_{\rm r}(K+Q)    (\Pi_{\rm hard}^{\lambda\sigma}(K+Q))_{\rm r}\,\Delta_{\rm r}(K+Q) \nonumber\bigg]\,,
\eea
where we use the notation ${\rm vtex}_{\mu\lambda\sigma\nu}$ to indicate the factor produced by contracting the two bare vertices and the metric tensors in the numerators of the gluon propagators in the Feynman gauge. 
Closing in the upper half plane so that only the poles of the advanced propagator contribute we make the replacements:
\bea
&& \Delta_{\rm a}(K) ~\to~ i\,{\rm Sgn}\,(k_0)\,\delta(K^2) \,, \\
&& \Delta_{\rm r}(K+Q) (\Pi_{\rm hard}^{\lambda\sigma}(K+Q))_{\rm r}\,\Delta_{\rm r}(K+Q) ~\to~ {\rm Re}\,\Pi_{\rm hard}^{\lambda\sigma}(K) \left(\frac{1}{k_0 q_0}\right)^2\,, \nonumber\\
&& N_b(k_0+q_0)-N_b(k_0) \to q_0 n_b^\prime(k_0)\,,\nonumber\\
&&{\rm vtex}_{\mu\lambda\sigma\nu}~\to~ {\rm vtex}_{\mu\lambda\sigma\nu}\bigg|_{Q=0}\,,\nonumber
\eea
which gives:
\bea
\label{integrand1z}
{\rm Re}\,\Pi_{\mu\nu}(q_0,0) \sim \frac{g^2}{q_0}\int dK\;&&{\rm vtex}_{\mu\lambda\sigma\nu}\bigg|_{Q=0}\; {\rm Sgn}\,(k_0)\,\delta(K^2)\,n_b^\prime(k_0)\,{\rm Re}\,\Pi_{\rm hard}^{\lambda\sigma}(K)\,\left(\frac{1}{k_0}\right)^2\,.\nonumber\\
\eea
This expression is of order $g\,({\rm LO})$ and does not have the right $q_0$ parity. 
The factor ${\rm vtex}_{\mu\lambda\sigma\nu}\bigg|_{Q=0}$ contains terms of the form
$
g_{\mu\lambda}K_\sigma K_\nu$, $K^2 g_{\mu\lambda} g_{\sigma\nu}$, $\cdots$ including all permutations of the indices. All terms with a factor $K^2$ are zero using the delta function, and terms with $K_\lambda$ or $K_\sigma$ are zero using the Ward identity. There will be a surviving term of the form $g_{\lambda\sigma}K_\mu K_\nu$. 
This term is zero for $\Pi_{\mu}^{~~\mu}(q_0,0)$ since the trace produces a factor $K^2$. For the components of the self-energy $\Pi_T(q_0,0)$, $\Pi_L(q_0,0)$ and $\Pi_{00}(q_0,0)$ the integrand is odd and gives zero when integrated over $k_0$. Using $\Pi$ to represent $\Pi_\mu^{~~\mu}$, $\Pi_L$, $\Pi_T$ or $\Pi_{00}$ the integrals for  ${\rm Re}\,\Pi(q_0,0)$ are of order $g^2\,({\rm LO}).$

The integral for the graph in figure \ref{gluon2loop}b is:
\bea
\label{integrand1w}
\Pi^{\rm r}_{\mu\nu}(q_0,0) \,\sim \,g^2\,i\int dK\;&&{\rm trce}\; \bigg[\bigg(N_f(k_0+q_0)-N_f(k_0)\bigg) \cdot\;\Delta_{\rm a}(K)\Delta^2_{\rm r}(K+Q) \nonumber\bigg]\,,\\[2mm]
&&{\rm trce} = {\rm Tr}\big[\gamma_\mu \Ksla \gamma_\nu (\Ksla+\Qsla) \,(\Sigma_{\rm hard}(K+Q))_{\rm r} \,(\Ksla+\Qsla) \big]\,.\nonumber
 \eea
Closing in the upper half plane and making the replacements:
\bea
&& \Delta_{\rm a}(K) ~\to~i \,{\rm Sgn}\,(k_0)\,\delta(K^2) \,, \\
&& \Delta^2_{\rm r}(K+Q) ~\to~  \left(\frac{1}{k_0 q_0}\right)^2\,, \nonumber\\
&& N_f(k_0+q_0)-N_f(k_0) \to q_0 n_f^\prime(k_0)\,,\nonumber\\
&&{\rm trce} ~\to~ {\rm trce}\bigg|_{Q=0}\,,\nonumber
\eea
we obtain: 
\bea
\label{integrand1v}
{\rm Re}\,\Pi_{\mu\nu}(q_0,0) \sim \frac{g^2}{q_0}\int dK\;&&{\rm trce}\bigg|_{Q=0}\; {\rm Sgn}\,(k_0)\,\delta(K^2)\,n_f^\prime(k_0)\,\left(\frac{1}{k_0}\right)^2\,,
\eea
which is of order $g\,({\rm LO})$ and does not have the right parity. 
The integrand is odd in $k_0$, not including the factor `trce,' which means that contributions to the trace that are even in $k_0$ will give zero when integrated. The trace contains terms with a factor $K^2$ which we set to zero and additional terms of the form:
\bea
\label{num}
 k_0 K_\mu K_\nu \cdot {\rm Re}\,\Sigma^{(0)}_{\rm hard}(K)\,,~~~~ k K_\mu K_\nu \cdot {\rm Re}\,\Sigma^{(s)}_{\rm hard}(K)\,.\nonumber
\eea
The component $\Pi_{\mu}^{~~\mu}(q_0,0)$ is zero since  the trace produces a factor $K^2$, and the components  $\Pi_T(q_0,0)$, $\Pi_L(q_0,0)$ and $\Pi_{00}(q_0,0)$ produce an even contribution to `trce' which gives zero when integrated. Thus the term of order $g({\rm LO})$ cancels and the result is $\sim\;g^2({\rm LO})$.
 
The conclusion is that for the real part all contributions to  2-loop propagator correction diagrams that are of order $g({\rm LO})$ cancel identically, and both propagator corrected and vertex corrected 2-loop diagrams contribute  at order $g^2({\rm LO})$. 

The imaginary parts of the 2-loop graphs are suppressed relative to the real parts by one factor of coupling. This is basically a consequence of the restriction of the phase space produced by the extra delta function. Note also that an extra factor $q_0/k_0$ is necessary to satisfy (\ref{parity}). We summarize these results for the (2-loop hard) contributions:
\bea
\label{2loopRes}
{\rm 2\!-\!loop~hard}:&&~~~~{\rm Re}\Sigma_{\rm 2lh} \sim g^2\,({\rm LO}) \sim g^3 T \,,~~~\\[2mm]
&& ~~~~{\rm Re}\Pi_{\rm 2lh} \sim g^2\,({\rm LO}) \sim g^4 T^2\,,\nonumber\\[2mm]
&& ~~~~{\rm Im}\Sigma_{\rm 2lh}  \sim g\,({\rm Re}\Sigma_{\rm 2lh}) \sim g^4 T \,,\nonumber\\[2mm]
&& ~~~~{\rm Im}\Pi_{\rm 2lh} \sim g\,({\rm Re}\Pi_{\rm 2lh})  \sim g^5 T^2 \nonumber\,.
\eea
These results for the real parts agree with what we found in the previous section for the (1-loop hard corrected)  contributions (see equation (\ref{1loopCorrected})). 

\subsection{(1-loop soft) diagrams}
\label{section:1s}
\begin{figure}[H]
\begin{center}
\includegraphics[width=16cm]{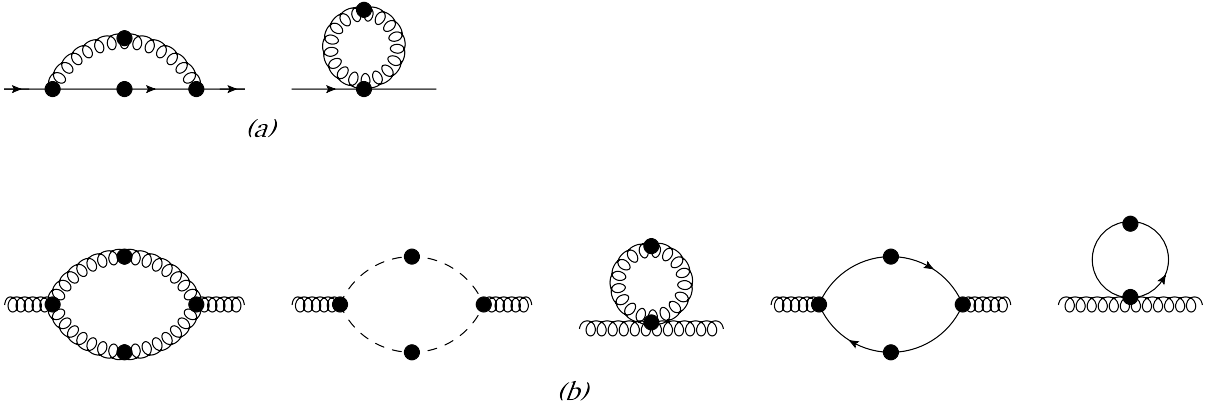}
\end{center}
\caption{(1-loop soft) contributions to the fermion and gluon self-energies. \label{1soft}  }
\end{figure}

Self-consistency requires that soft lines, and vertices with all legs soft, be replaced by HTL resummed lines and vertices. Thus the (1-loop soft) contribution comes from the diagrams in figures \ref{1soft}a and \ref{1soft}b for the fermion and gluon self-energies. In QED the 3-photon vertex is identically zero by Furry's theorem, and the HTL 4-photon vertex is also zero because it is traceless and proportional to the 3-photon vertex. For the photon self-energy therefore, we need only the QED analogue of the last two diagrams in figure \ref{1soft}b. 

We start with some general comments on the order of the imaginary part relative to the real part. 
In section \ref{section:2h} we have seen that, for hard integrals, the imaginary part is of order $(q_0/T\sim g)$ relative to the real part. The suppression of the imaginary part occurs because the integrand for the imaginary part contains an extra delta function and, for hard loop momenta, this delta function restricts the phase space and reduces the order of the integral. Note also that the factor $q_0/T$ is necessary to change the parity of the imaginary part relative to the real part, so that  (\ref{parity}) is satisfied. 
For the (1-loop soft) contributions, the extra delta function does not restrict the phase space (because the integral is over soft momenta) and one can change the parity of the imaginary part relative to the real part without changing the order of the integral,  because there is an extra momentum scale in a calculation involving resummed lines and vertices.
Using resummed propagators, one obtains expressions that depend on a thermal mass (from the real part of the LO self-energy) and the imaginary part is of order $(q_0/m_{\rm th}\sim 1)$ relative to the real part. Thus it has opposite parity but is the same order.

To determine the order of the (1-loop soft) diagrams in figure \ref{1soft}, we look at the corresponding graphs with bare lines and vertices. This will give the correct order of the 1-loop soft contribution (up to logarithmic factors), even though it does not give the correct result. 
Since both internal and external momenta are soft, this amounts to using dimensional analysis, modified by the appropriate factor from the distribution functions. The integrand for a 1-loop diagram contains a sum of terms with one symmetric propagator on each line (see figure \ref{symAll}). Using equations (\ref{defn:distro}) and ({\ref{keldyshProp1}) we need to look at the factors $N_b(p_0)$ and $N_f(p_0)$ in the limit $T\gg p_0$. Taking the limit gives:
\begin{eqnarray}
\label{expand-distro}
&&  N_b(p_0) = 1+\frac{2}{e^{p_0 \beta}-1} ~~\to~~ \frac{T}{p_0} \sim \frac{1}{g}\,,\\
&&  N_f(p_0) = 1-\frac{2}{e^{p_0 \beta}+1} ~~\to~~\frac{p_0}{T} \sim g\,.\nonumber
\end{eqnarray}
Both diagrams in the fermion self-energy have a soft boson on an internal line, and therefore both will have at least one term with a $1/g$ Bose-Einstein enhancement factor. The first three diagrams in figure \ref{1soft}b contain internal soft bosons, so the gluon self-energy will be dominated by these graphs and also have a factor $1/g$ from the distribution functions. The last two graphs, whose QED analogues produce the photon self-energy, get a factor $\sim g$ from the distribution functions. 

Combining these results we have:
\bea
\label{bareSoftFerm}
{\rm 1~loop~bare~soft}:&&~~~\Sigma  \sim g^2 (gT)\frac{1}{g} \sim g^2 T  \sim g\,({\rm LO})\,,\\
\label{bareSoftGluon}
&&~~~\Pi_{\rm gluon}\sim g^2 (gT)^2\frac{1}{g} \sim g^3 T^2 \sim g\,({\rm LO})\,,\\[2mm]
\label{bareSoftPhoton}
&& ~~~\Pi_{\rm photon}\sim e^2 (eT)^2 \;e\sim e^5 T^2 \sim e^3\,({\rm LO})\,.
\eea
In each of these expressions, the factor (coupling constant)$^2$ comes from the vertices, the factor in brackets is  from the dimensions of the diagram together with the fact that all momenta are assumed soft, and the last factor is from the distribution functions. 

Comparing the results of equations (\ref{1loopCorrected}), (\ref{2loopRes}), (\ref{bareSoftFerm}), (\ref{bareSoftGluon}) and (\ref{bareSoftPhoton}) we conclude that for gluons and fermions, the full NLO contribution to the self-energy comes from the (1-loop soft) contribution. This result was verified by brute force in Refs. \cite{Schulz1994} (for gluons) and \cite{Zsolt} (for fermions). For photons the situation is more complicated. This is the subject of the next section. 

\section{Photon self-energy}
\label{section:photons}

The imaginary part of the photon self-energy is related to rates of scattering and production processes and is particularly interesting in the context of quark-gluon plasma. From equations (\ref{2loopRes}) and (\ref{bareSoftPhoton}), it appears that the imaginary part of the photon self-energy is the only 2-point function for which one must calculate both the (1-loop soft) and (2-loop hard) pieces to obtain the full NLO result. The photon self-energy is worthy of careful study because we would like to know if it is unique in this regard, or if there are other  $n$-point functions which also receive contributions from the (2-loop hard) sector. In addition, we have so far ignored logarithmic factors, and therefore it is possible that either the (1-loop soft) or (2-loop hard) pieces might be logarithmically enhanced over the other. 
The calculation of the (1-loop soft) piece was done by Braaten, Pisarski and Yuan \cite{BPY}. The fact that the (2-loop hard) part is necessary was recognized almost 10 years later by Aurenche et al. \cite{PAbrem,PAcomp}.  In this section we discuss these results from a power counting point of view, as a first step towards understanding the behaviour of higher $n$-point functions.

The dressed propagators in the (1-loop soft) diagrams contain both pole and cut terms which represent different contributions to a given physical process.  
 The spectral functions for the HTL fermion propagators are:
\bea
&&\rho_{\pm}(p_0,p) = 2\pi \big[Z_\pm(p)\delta(p_0-\omega_{\pm}(p))+Z_\mp(p)\delta(p_0+\omega_{\mp}(p))\big]+\beta(p_0,p)\,,\\[2mm]
&& Z_\pm (p) = \frac{\omega_\pm^2(p)-p^2}{2m_f^2} \,,\nonumber\\
&&\beta(p_0,p) = \theta\left(p^2-p_0^2\right)\frac{\pi  m_f^2 \left(1\mp \frac{p_0}{p}\right)\frac{1}{p}}
{
\left(p \left(1\mp \frac{p_0}{p}\right)
\pm \frac{m_f^2}{2p} \left[\left(1\mp\frac{p_0}{p}\right) \ln
   \left(\frac{p_0+p}{p_0-p}\right)\pm 2 \right]\right)^2 
   + \frac{\pi^2 m_f^4}{4 p^2} \left(1\mp\frac{p_0}{p}\right)^2}\,.\nonumber
\eea
The term containing the delta function is called the `pole' part, and the term with the theta function is the `cut' part. The 1-loop bubble diagram contains the product of two spectral functions and therefore has three different kinds of terms which we call pole-pole, pole-cut and cut-cut. 
The pole-pole term is a kind of modified the Born term, the pole-cut terms contain 2-loop effects like Compton scattering and particle annihilation, and the cut-cut piece includes 3-loop processes like bremsstrahlung and off-shell annihilation.
 
Mathematically, cuts contain an extra factor~ $\sim\,(m_f^2/P^2)$ relative to poles and therefore for $P\sim eT$ it appears that poles and cuts contribute at the same order. However, this ignores the possible appearance of logarithmic factors, which we have not yet considered. 
Power-counting arguments, from their very nature, depend on the separation of momenta into soft and hard scales. Since a logarithm is produced by an integral of the form: 
\bea
\label{logForm}
\int_{p_{\rm soft}}^{p_{\rm hard}}dp\;\frac{1}{p} \sim {\rm ln}\frac{1}{g}
\eea
which samples the full range of the momentum integration, 
it is clear that logarithms will necessarily be missed if it is assumed that all momenta are either hard or soft. Our previous analysis amounted to the approximation $\int dp\;1/p \sim 1$. 

In order to determine if a logarithmic enhancement occurs, we must look more carefully at the structure of the integrand. 
We consider (1-loop soft) diagrams and try to see how a logarithm could appear. It is clear that pole-pole terms will not produce a log, since the product of delta functions restricts both the $p_0$ and $p$ integrals to soft momenta. For terms that contain cuts, at least one delta function is missing and therefore at least one integral will be restricted only by a theta function. To see how a logarithm could appear we assume that the integral is dominated by the region of phase space for which the loop momentum $P$ is softer than $T$ but harder than the other soft scales in the integral. We call a momentum like this soft$^+$ and denote it $P^+$.
We expand the integrand in $(q_0/P^+)$. 
If the integrals produced by expansion are less than logarithmically divergent, then the dominant piece of the integral comes from $P^+$ not soft$^+$ but a true hard scale, and the 1-loop diagram is just the HTL. 
If the integral is more than logarithmically divergent, then it is dominated by $P^+\ll q_0$ and $P^+$ is not soft$^+$ but true soft, which means the integral should not be expanded in $(q_0/P^+)$. In this case, the order of the integral is correctly predicted by equations (\ref{bareSoftFerm}) - (\ref{bareSoftPhoton}). 
If the expansion gives a leading term of order $(q_0/P^+)$ the $P$ integral produces a logarithmic factor.
The result of the calculation \cite{BPY,PAcomp,PAbrem} is that the pole-pole piece is of order $e^5 T^2$ (in agreement with (\ref{bareSoftPhoton})) but terms which contain cuts are of order  $e^5{\rm ln}(1/e)\,T^2$. Terms with cuts also have more singular behaviour as $q_0\to 0$ and can dominate the pole-pole contribution for $q_0$ small.


Next we consider how we could get a log in a (2-loop hard) diagram. 
We assume the integral is dominated by the part of phase space for which one loop momentum is softer than the temperature and the other loop momentum, but still harder than the soft scale $eT$. This momentum will be called hard$^-$ and denoted $L^-$, the other loop momentum is $K$.
 We expand in $(L^-/K)$. If the $L$ integral is more than logarithmically divergent, then the integral is dominated by $L$ not hard$^-$ but a true soft scale. In this case the leading term in the expansion reproduces a piece of the (1-loop soft) contribution, with the integration over $K$ giving an HTL vertex or propagator insertion. 
If the $L$ integral is less than logarithmically divergent the dominant contribution comes from $K \sim L^-$, which means $L^-$ is not hard$^-$ and the integral  should not be expanded in $(L^-/K)$.
If the expansion gives a leading term of order $(L^-/K)^{-1}$ the $L$ integral produces a logarithmic factor. The result of the calculation in Refs. \cite{PAbrem,PAcomp} is that if one chooses the hard$^-$ momentum on the internal boson line (see figure \ref{fig:tadpole}) 
there is a logarithmic factor which modifies (\ref{2loopRes}) and gives (2-loop hard) diagrams of order $\sim\,e^5 \ln 1/e\,T^2$. Thus the (1-loop soft) and (2-loop hard) pieces contribute at the same order.

The type of integrals that produce a log from soft$^+$ and hard$^-$ scales are shown symbolically in equation (\ref{log-soft-hard}). 
\bea
\label{log-soft-hard}
&& \int_{p_{\rm soft}}^{p_{\rm hard}}\frac{dP}{P+Q} \sim \int_{p_{\rm soft}}^{p_{\rm hard}}\frac{dP}{P} = {\rm ln}\frac{p_{\rm hard}}{p_{\rm soft}}= {\rm ln}\frac{1}{e} ~~ {\rm for} ~~  q_0 < P^+ <T\,. \\[2mm]
&& \int_{l_{\rm soft}}^{l_{\rm hard}}\frac{dL}{L+L^2/K} \sim \int_{l_{\rm soft}}^{l_{\rm hard}}\frac{dL}{L} \sim {\rm ln}\frac{l_{\rm hard}}{l_{\rm soft}}\sim {\rm ln}\frac{1}{e} ~~ {\rm for} ~~   q_0 < L^- < K\sim T\,.\nonumber
\eea

Notice however that if we include (1-loop soft) diagrams with soft$^+$ momenta and (2-loop hard) diagrams with one momentum hard$^-$ there will be an overlap in the regions of phase space which produce these two contributions. Thus we would expect that the (2-loop hard) contribution with one momentum hard$^{-}$ is already included in the (1-loop soft) piece with momentum soft$^+$. The 
HTL effective Lagrangian is constructed with counter-terms that subtract order by order the loop corrections already included in the effective Lagrangian, and thus the 2-loop diagrams which contain 1-loop counter-terms should exactly remove the NLO contribution from the 2-loop diagrams, which would mean that once again (for a different reason), the (2-loop hard) contribution does not need to be calculated. 

However, this argument fails for the photon self-energy: the (2-loop hard) piece is not included in the (1-loop soft) part and must be calculated separately.  In order to know how to proceed at higher orders, we need to understand what is wrong with the argument above in the case of the photon self-energy. 

To determine if the (1-loop soft) contribution contains everything in the (2-loop hard) piece, we could generate a series of perturbative diagrams by expanding the (1-loop soft) diagrams in $e$ (holding $q_0$ fixed). If all 2-loop diagrams are present in the expansion of the (1-loop soft) diagrams, then we do not need the (2-loop hard) contribution, but if some diagrams are missing in the perturbative expansion, then we need to look explicitly the corresponding terms from the (2-loop hard) part. 


There is another very simple way to see that 2-loop diagrams are needed, and also what region of phase space is important. As discussed in section \ref{section:1s}, the (1-loop soft) photon self-energy does not contain a photon tadpole, since the 4-photon vertex is zero in the HTL approximation. This result is obtained from the fact the 4-photon vertex is traceless in LO HTL and contractions are proportional to the 3-photon vertex, which is zero by Furry's theorem. 
However, if we consider photon legs that are soft$^+$ instead of soft, we would obtain a sub-leading logarithmic contribution to the HTL 4-photon vertex which we call the `HTL-extended' 4-photon vertex. If soft$^+$ momenta dominate, the photon tadpole diagram constructed with an HTL-extended 4-photon vertex should be included in the (1-loop soft) calculation. From equation (\ref{LOorder}), the order of the HTL 4-photon vertex would be $e^2$  (if it were not zero using Furry's theorem). 
If one of the legs is soft$^+$, we expect a non-zero contribution of order $e^2(e\,{\rm ln}1/e)$. Multiplying by a factor $(eT)^2$ from the dimensions of the self-energy, we expect the tadpole diagram with an HTL-extended 4-photon vertex to be of order $e^5\,{\rm ln}1/e\,T^2$. 

If we drop this tadpole diagram, we will miss terms in the 2-loop expansion which must then be included explicitly as part of the (2-loop hard) piece. Equivalently, the soft$^+$ tadpole diagram reappears as the hard$^-$ 2-loop diagrams with the momentum of the fermion loop $K$ taken to be hard, and the momentum of the gauge boson line $L$ taken hard$^-$. The authors of Refs. \cite{PAbrem,PAcomp}  found that the leading piece of the (2-loop hard) contribution has precisely this origin and is of order $e^5\,{\rm ln}1/e\,T^2$. 
The correspondence of these two pieces is clear at a diagrammatic level and is shown in figure \ref{fig:tadpole}. 
\begin{figure}[H]
\begin{center}
\includegraphics[width=18cm]{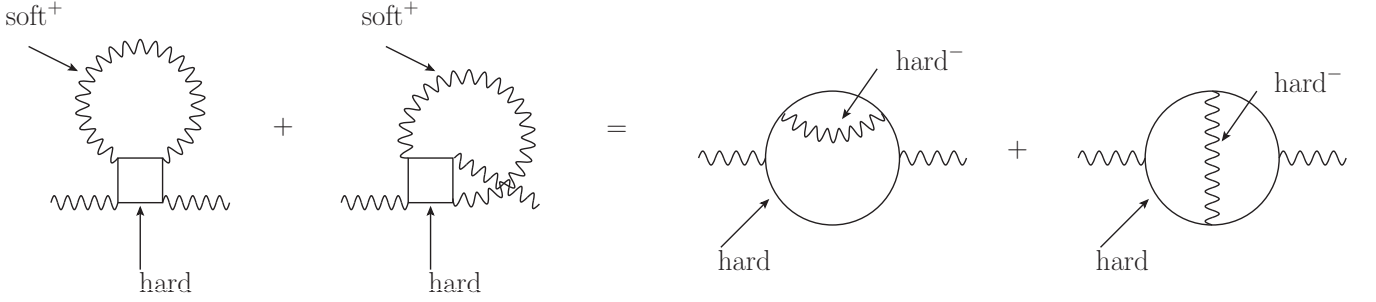}
\end{center}
\caption{Two different ways to represent the same NLO contribution to the photon self-energy. The second graph represents the contribution from the crossed-box diagram in the HTL extended 4-photon vertex.  \label{fig:tadpole}  }
\end{figure}

\section{Summary of Results}
\label{section:table}

In the table below we summarize the results for the NLO 2-point functions from QCD and QED. For quantities which have not been calculated, the entries in the table come from the power-counting results in equations (\ref{1loopCorrected}), (\ref{2loopRes}) and (\ref{bareSoftPhoton}}) and do not include possible logarithmic factors. Some of the sub-leading contributions which were calculated may also have logarithmic factors that were dropped.  \\

\begin{center}

\begin{tabular}{|l|l|l r |l r  |}  \hline
Quantity & Term &  Re~~~~~~~~~~& Ref &  Im~~~~~~~~~~ & Ref  \\  \hline
$\Pi_g$ & LO & $g^2 T^2$ &\cite{BP} &  0& \cite{BP}\\ \hline
		& 1-loop soft & $g^3 T^2$ & \cite{Schulz1994} &  $g^3 T^2$ & \cite{BPgluon} \\ \hline
		& 2-loop hard & $g^4 T^2$ &  \cite{Schulz1994} & $g^5 T^2$ &  \\ \hline
		& 1-loop hard correction & $g^4 T^2$  & \cite{Schulz1994} & - &  \\ \hline\hline
$\Pi_\gamma$ & LO & $e^2 T^2$ &\cite{BP} &  0& \cite{BP}\\ \hline
		& 1-loop soft & $e^5 T^2$ & &  $e^5 {\rm ln}\frac{1}{e} T^2$ & \cite{BPY,PAbrem,PAcomp} \\ \hline
		& 2-loop hard & $e^4 T^2$ &   & $e^5{\rm ln}\frac{1}{e} T^2$ & \cite{PAbrem,PAcomp} \\ \hline
		& 1-loop hard correction & $e^4 T^2$ &  & - &  \\ \hline\hline
$\Sigma$ & LO & $g T$ &\cite{BP} &  0& \cite{BP}\\ \hline
		& 1-loop soft & $g^2 T$ & \cite{Carrington2008}&  $g^2T$ & \cite{BPquark,KKM,Carrington2006} \\ \hline
		& 2-loop hard & $g^3 {\rm ln}\frac{1}{g} T$ &  \cite{Emil} & $g^4 T$ & \cite{Zsolt} \\ \hline
		& 1-loop hard correction & $g^3 {\rm ln}\frac{1}{g} T$ & \cite{Mitra} & - &  \\ \hline\hline
		
		\end{tabular}

\end{center}

\vspace*{.2cm}

We would like to know if there are other $n$-point functions, besides the photon self-energy, which also receive NLO contributions from the (2-loop hard) sector that are not part of the (1-loop soft) piece. 

First we review what conditions conspire to create the situation where the (2-loop hard) diagrams need to be calculated to obtain the imaginary part of the photon self-energy at NLO. There are two important points.
(1) The (1-loop soft) part for photons is suppressed relative to fermions and gluons (because all internal lines are fermions) and is the same order (up to logs) as the 2-loop hard piece. (2) Both the (1-loop soft) and (2-loop hard) pieces have a logarithmic enhancement, which means there is an overlap in the phase spaces of the two contributions, but there is also a missing tadpole, which means counter-terms don't remove the (2-loop hard) part. 

For higher $n$-point functions, the only vertices for which all internal lines are fermions are $2n$-photon vertices. 
Thus we expect that the (1-loop soft) and (2-loop hard) sectors can be the same order, up to logs, for a $2n$-photon vertex and no other vertices. 

\begin{figure}[H]
\begin{center}
\includegraphics[width=8cm]{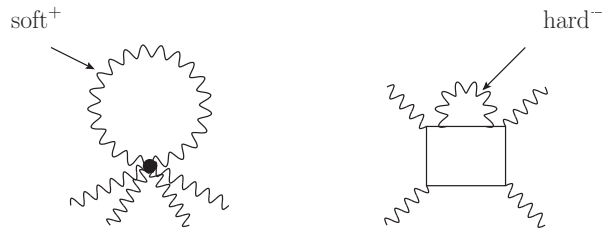}
\end{center}
\caption{The contribution of a tadpole diagram built from an HTL-extended 6-point vertex,  and a hard$^-$ 2-loop graph to the NLO 4-photon vertex. \label{fig:4photon}  }
\end{figure}

If either the (1-loop soft) or (2-loop hard) piece has a log enhancement, the other is not needed. 
If neither piece has a log, there is no overlap in the phase spaces and the counter-terms cannot remove the (2-loop hard) piece - both must be calculated. 
If both pieces have logs, there is an overlap in the regions of phase space that contribute to the two pieces, and in principle the counter-terms should completely remove the (2-loop hard) piece.  However, if there is a tadpole `missing' from the (1-loop soft) contribution, the corresponding hard$^-$ part of the relevant (2-loop hard) diagram would have to be included. 
Missing tadpoles correspond to HTL vertices that are zero when all legs are soft, but non-zero if some leg momenta are soft$^+$. 
All photon-only vertices are zero in LO HTL, since the odd ones are zero by Furry's theorem and the even ones can be written as a sum of terms proportional to traces and lower order odd vertices. A $2n$-photon vertex with some legs soft$^+$ will not be identically zero but will have a sub-leading correction. 
This means that for any $2n$-photon vertex the tadpole diagram with a $2(n+1)$-photon HTL vertex will be missing from the (1-loop soft) piece. 
For example, the 4-photon vertex would get a logarithmically enhanced NLO contribution from the missing tadpole in figure \ref{fig:4photon}a, or equivalently from the (2-loop hard) diagram in figure \ref{fig:4photon}b. 

The conclusion is that we expect that the only cases for which one does not get the full NLO result from the (1-loop soft) piece are $2n$-photon vertices. 

\section{Conclusions}
\label{section:conclusions}
The hard-thermal-loop effective theory has been around for over 20 years, but very little progress has been made beyond leading order. Part of the problem is that standard power-counting based on the imaginary time formalism predicts that a large number of diagrams may need to be calculated, including soft 1-loop diagrams with effective lines and vertices, and some 2-loop diagrams with hard loop momenta (which we have called (1-loop soft) and (2-loop hard) contributions).
However, with the exception of the photon self-energy, in all of the calculations that have been done the full next-to-leading order contribution can be obtained by calculating only (1-loop soft) diagrams. In this paper we have performed a refined power-counting analysis using the Keldysh representation of real-time finite temperature field theory. 
We have shown that the contribution of (2-loop hard) diagrams is over estimated by previous power counting rules, which explains why the (2-loop hard) diagrams do not contribute at NLO to the fermion and gluon self-energy. 
The (2-loop hard) diagrams do contribute to the photon self-energy because: (1) the (1-loop soft) piece has an extra suppression factor which make it the same order as the (2-loop hard) part and; (2) the (2-loop hard) piece is not removed by the HTL counter-terms. 
For higher $n$-point functions the same situation can occur for $2n$-photon vertices if both pieces have logs or if both pieces do  not have logs, for different reasons. If both pieces do not have logs, then there is no overlap in phase space and the  counter-terms cannot remove the contribution of the (2-loop hard) diagrams. 
If both pieces do have logs, then there is overlap, but the piece of phase space that produces the log is exactly the piece that gives a missing HTL-extended tadpole diagram, which can alternatively be included as the corresponding (2-loop hard$^-$) diagram. 

Based on these arguments, we expect that the only possible exceptions to the rule that the 1-loop soft diagrams give the full next-to-leading order contribution are $2n$-photon vertices.

\end{document}